\documentclass[prd,onecolumn,
  tightenlines,nofootinbib,eqsecnum,
]{revtex4}

\usepackage[francais]{babel}

\usepackage{amsmath}
\usepackage{amsfonts}
\usepackage{amssymb}
\usepackage{bm}
\usepackage{hyperref}
\usepackage{mathrsfs}
\usepackage{graphicx}

\usepackage{empheq}

\usepackage{ulem}
\usepackage[usenames]{color}

\definecolor{darkgreen}{rgb}{0,0.5,0}

\hypersetup{
    bookmarks=true,         
    unicode=false,          
    pdftoolbar=true,        
    pdfmenubar=true,        
    pdffitwindow=false,     
    pdfstartview={FitH},    
    pdftitle={My title},    
    pdfauthor={Author},     
    pdfsubject={Subject},   
    pdfcreator={Creator},   
    pdfproducer={Producer}, 
    pdfkeywords={keyword1} {key2} {key3}, 
    pdfnewwindow=true,      
    colorlinks=true,       
    linkcolor=red,          
    citecolor=cyan,        
    filecolor=magenta,      
    urlcolor=darkgreen,           
    linktocpage=true
}

\allowdisplaybreaks

\DeclareSymbolFontAlphabet{\mathrsfs}{rsfs}
\DeclareMathAlphabet{\mathcal}{OMS}{cmsy}{m}{n}

\newcommand{\beq}{\begin{equation}}
\newcommand{\eeq}{\end{equation}} 

\newcommand{\ud}{\mathrm{d}}

\begin{document}

\title{Les Ondes Gravitationnelles 100 ans apr\`es
  Einstein\\{\small{\rm [\`A para\^itre dans ``\textit{Reflets de la
          Physique}'', revue de la Soci\'et\'e Fran\c{c}aise de
        Physique (2017)]}}}


\author{Luc \textsc{Blanchet}}\email{luc.blanchet@iap.fr}
\affiliation{Institut d'Astrophysique de Paris, UMR 7095 du CNRS,
  Sorbonne Universit{\'e}s \& UPMC Universit\'e Paris
  6,\\ 98\textsuperscript{bis} boulevard Arago, 75014 Paris, France}

\date{\today}

\begin{abstract}
La collaboration LIGO-VIRGO a d\'etect\'e directement sur Terre le
passage d'ondes gravitationnelles \'emises lors de la collision et de
la fusion de deux trous noirs massifs \`a une distance
astronomique. Cette d\'ecouverte majeure ouvre la voie \`a
l'Astronomie gravitationnelle, qui devrait r\'evolutionner notre
connaissance de la structure de l'Univers aux grandes \'echelles, avec
notamment les m\'ecanismes de formation des trous noirs et leur r\^ole
dans l'\'evolution de l'Univers, l'\'emergence probable d'une
astronomie ``multi-messag\`ere'' conjointe avec le rayonnement
\'electromagn\'etique, ainsi qu'une meilleure appr\'ehension de la
place de la relativit\'e g\'en\'erale par rapport aux autres
interactions fondamentales. Les travaux th\'eoriques et num\'eriques
sur le probl\`eme \`a deux corps en relativit\'e g\'en\'erale jouent
un r\^ole tr\`es important dans le d\'echiffrage et l'interpr\'etation
des signaux d'ondes gravitationnelles.
\end{abstract}


\maketitle

\vspace{0.3cm}
\section{Qu'est-ce qu'une onde gravitationnelle\,?} 
\vspace{0.3cm}

Les ondes gravitationnelles sont pr\'edites par toute th\'eorie de la
gravitation ``relativiste'', c'est-\`a-dire en accord avec les
principes de la relativit\'e restreinte de Einstein, Lorentz et
Poincar\'e de 1905. Ainsi, d\`es 1906, dans le cadre de sa th\'eorie
relativiste de la gravitation, Poincar\'e avait propos\'e le concept
d'``ondes gravifiques''. Mais c'est bien s\^ur \`a Einstein, dans son
article de 1916, soit quelques mois apr\`es la publication de la
relativit\'e g\'en\'erale en novembre 1915, que revient la formulation
moderne des ondes gravitationnelles.

Dans le cadre de la relativit\'e g\'en\'erale, une onde
gravitationnelle est une d\'eformation de la g\'eom\'etrie de
l'espace-temps se propageant \`a la vitesse de la lumi\`ere. Elle est
produite dans des zones de champs gravitationnels intenses, par des
sources form\'ees d'\'enormes quantit\'es de mati\`ere se
d\'epla\c{c}ant \`a des vitesses proches de celle de la
lumi\`ere. C'est le mouvement coh\'erent d'ensemble de la mati\`ere
qui engendre l'onde gravitationnelle, au contraire du rayonnement
\'electromagn\'etique qui est en g\'en\'eral produit par la
superposition des \'emissions individuelles des atomes ou mol\'ecules
constituant la source. A cet \'egard il existe une analogie profonde
entre une onde gravitationnelle et une onde sonore. Comme pour l'onde
sonore, mais au contraire d'une onde \'electromagn\'etique, l'onde
gravitationnelle a une longueur d'onde tr\`es grande par rapport \`a
la taille de sa source. Cette diff\'erence essentielle avec les ondes
\'electromagn\'etiques rend l'astronomie des ondes gravitationnelles
tr\`es diff\'erente de l'astronomie traditionnelle.

Alors que les ondes \'electromagn\'etiques sont ais\'ement d\'evi\'ees
et absorb\'ees par la mati\`ere, le rayonnement gravitationnel
interagit extr\^emement faiblement avec la mati\`ere, et peut donc se
propager sans impunit\'e sur des tr\`es grandes distances. En
particulier, il n'existe pas d'``\'ecran'' \`a une onde
gravitationnelle, car la masse, qui est la ``charge'' associ\'ee \`a la
force gravitationnelle, est toujours positive. Par contre le
rayonnement gravitationnel subit comme la lumi\`ere des d\'eviations
par les champs gravitationnels, produits par exemple par les grandes
structures dans l'Univers.

En principe on peut associer \`a l'onde gravitationnelle une particule
qu'on appelle le graviton, de la m\^eme fa\c{c}on que le photon est
associ\'e \`a l'onde \'electromagn\'etique. Comme le photon, le
graviton a une masse nulle et deux \'etats de polarisation. Par contre
il a un spin \'egal \`a deux, double de celui du photon, ce qui est
li\'e \`a la nature tensorielle du champ gravitationnel en
relativit\'e g\'en\'erale. Si on essaye d'imaginer des ordres de
grandeurs pour les probabilit\'es de processus quantiques d'\'emission
ou d'absorption de gravitons, on trouve qu'elles sont extr\^emement
faibles, et il n'y a gu\`ere d'espoir que l'on puisse un jour observer
une transition ``quantique gravitationnelle''. Les ondes
gravitationnelles que l'on d\'etecte sont des ondes extr\^emement
``classiques''. De plus il n'existe malheureusement pas \`a l'heure
actuelle de th\'eorie quantique de la gravitation satisfaisante qui
permettrait d'unifier la relativit\'e g\'en\'erale et les autres
interactions connues.

\vspace{0.3cm}
\section{La r\'ealit\'e physique des ondes gravitationnelles}
\vspace{0.3cm}

Dans son article de 1918~\cite{E18}, Einstein calcule le flux
d'\'energie total \'emis sous forme d'ondes gravitationnelles par un
syst\`eme de mati\`ere, dans l'approximation dominante o\`u l'on
n\'eglige les corrections relativistes, et obtient ce qu'on appelle
aujourd'hui la formule du quadrup\^ole d'Einstein.\footnote{Einstein
  commet dans cet article une fameuse erreur de calcul\,: son
  r\'esultat est faux par un facteur 2\,!}  Une am\'elioration
cruciale de cette formule a \'et\'e apport\'ee par Landau et Lifchitz
dans leur trait\'e de Th\'eorie des Champs de 1947, o\`u ils montrent
que la formule est valable dans le cas d'un syst\`eme de mati\`ere
auto-gravitant, c'est-\`a-dire dont les mouvements sont engendr\'es
par la force gravitationnelle elle-m\^eme. La formule s'applique donc
\`a un syst\`eme binaire d'\'etoiles en mouvement sous l'action de la
force gravitationnelle newtonienne.

Avec les connaissances actuelles (notamment sur les m\'ethodes
d'approximation en relativit\'e g\'en\'erale), la formule du
quadrup\^ole nous para\^it suffisante pour \^etre convaincu de la
r\'ealit\'e physique des ondes gravitationnelles. Mais un d\'ebat,
apparaissant maintenant comme quelque peu surr\'ealiste, a longtemps
fait rage concernant leur existence. Eddington \'etait connu pour
\^etre tr\`es sceptique (notamment dans son livre de 1922), et
Einstein lui-m\^eme douta par deux fois du rayonnement gravitationnel
avant de se r\'etracter. Il n'a en tous cas jamais envisag\'e la
d\'etection directe des ondes gravitationnelles, et il ne croyait pas
non plus que les trous noirs existent r\'eellement. Ainsi, dans son
travail avec Rosen de 1937, il conclut que \guillemotleft\textit{les
  ondes gravitationnelles n'existent pas, quoiqu'elles aient \'et\'e
  consid\'er\'ees comme une certitude en premi\`ere
  approximation}\guillemotright, mais ce travail \'etait entach\'e
d'une erreur, comme l'a fait remarquer le rapporteur de son article
dans la Physical Review.\footnote{L'article d'Eintein fut rejet\'e par
  la Physical Review et publi\'e plus tard (avec la r\'etractation
  d'Einstein) dans une revue plus obscure. Le nom du rapporteur est
  longtemps rest\'e inconnu, mais les historiens des sciences ont
  d\'etermin\'e qu'il s'agissait de Robertson, le cosmologiste de la
  m\'etrique de ``Friedmann-Lema\^itre-Robertson-Walker''.}
Finalement, dans les ann\'ees 1960, des travaux de Bondi, Sachs et
Penrose sur la structure du champ gravitationnel loin de la source ont
\'etabli dans un cadre rigoureux que les ondes gravitationnelles
transportent de l'\'energie, qui est extraite de la masse de la
source, et peut \^etre d\'epos\'ee sur un d\'etecteur. En principe le
d\'ebat \'etait clos\,!  Mais m\^eme dans les ann\'ees 1970, la
confusion r\'egnait encore sur le probl\`eme de la ``r\'eaction de
rayonnement'', \`a savoir quel est l'impact de l'\'emission du
rayonnement gravitationnel sur le mouvement de la source.

La premi\`ere mise en \'evidence des ondes gravitationnelles remonte
\`a la fin des ann\'ees 1960 avec des observations de syst\`emes
d'\'etoiles binaires dits ``cataclysmiques'', constitu\'es d'une
\'etoile normale en fin de vie, en orbite autour d'une \'etoile
compacte, en l'occurence une naine blanche. L'\'etoile normale
d\'everse de la mati\`ere sur l'\'etoile compacte, qui forme un disque
d'accr\'etion o\`u cette mati\`ere est chauff\'ee par son mouvement
dans le fort champ gravitationnel. On observe ces syst\`emes gr\^ace aux
rayonnements UV et X \'emis par le disque d'accr\'etion. Il se trouve
que dans certains cas, pour des p\'eriodes orbitales tr\`es courtes,
ces syst\`emes binaires ne sont stables que si l'on invoque une perte
de moment cin\'etique orbital, et que seule l'\'emission d'ondes
gravitationnelles permet d'emporter du moment cin\'etique et
d'expliquer la stabilit\'e observ\'ee. Il est assez ironique de penser
que cette v\'erification fut faite par les astronomes (qui ne mettent
pas en doute les ondes gravitationnelles) \`a une \'epoque o\`u
certains physiciens relativistes \'etaient emp\^etr\'es dans leurs
controverses\,!

\begin{figure}[t]
\begin{center}
\includegraphics[width=12cm]{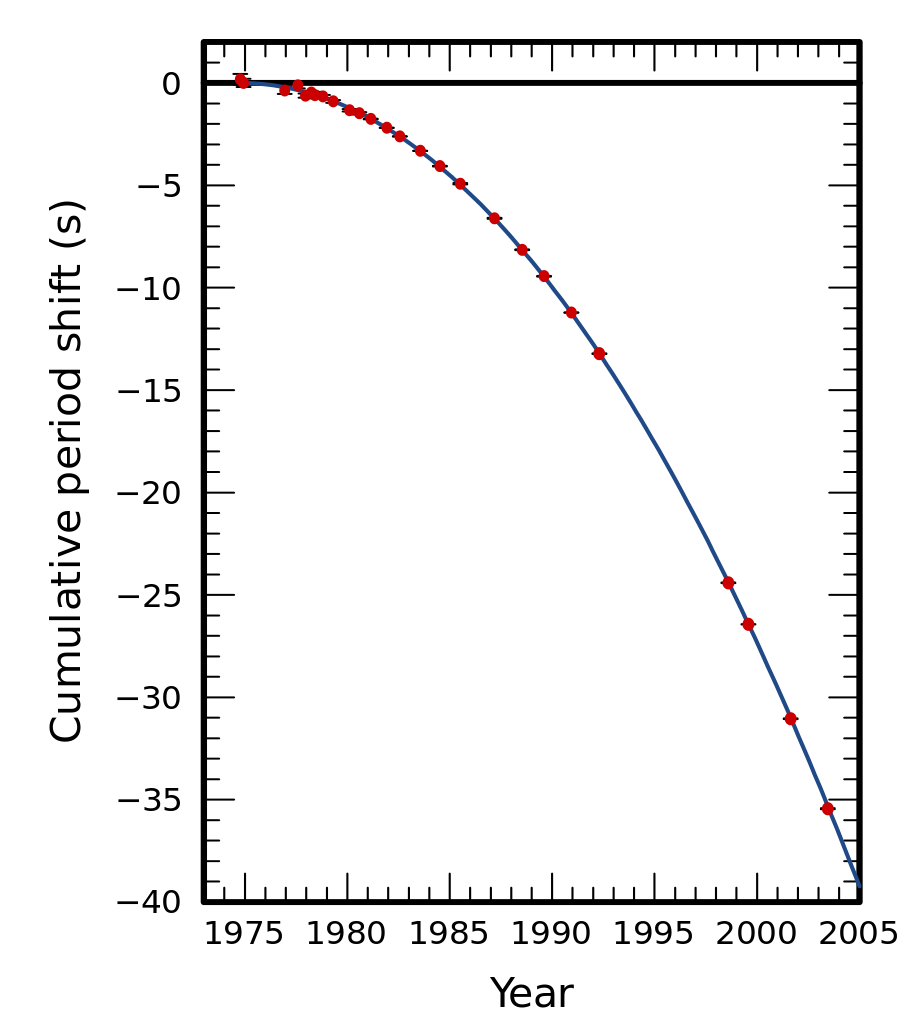}
\end{center}
\caption{La d\'ecroissance de la p\'eriode orbitale du pulsar binaire
  PSR B1913+16~\cite{TW82}. Les observations (points rouges) sont
  parfaitement conformes avec la pr\'ediction de la relativit\'e
  g\'en\'erale (ligne bleue) et, en fait, de la formule du
  quadrup\^ole d'Einstein. Hulse et Taylor ont re\c{c}u le prix Nobel
  en 1993 \guillemotleft\textit{pour la d\'ecouverte d'un nouveau type
    de pulsar, une d\'ecouverte qui a ouvert des nouvelles
    possibilit\'es pour l'\'etude de la
    gravitation}\guillemotright.}\label{fig1}
\end{figure}

La premi\`ere preuve quantitativement pr\'ecise des ondes
gravitationnelles fut obtenue gr\^ace au pulsar binaire d\'ecouvert par
Hulse et Taylor en 1974~\cite{HulseTaylor}. Un pulsar est une \'etoile
\`a neutrons fortement magn\'etis\'ee et en rotation rapide, que l'on
observe par les pulses radio \'emis le long de l'axe magn\'etique \`a
chaque rotation en direction de la Terre. Ce pulsar est en orbite
rapproch\'ee autour d'une autre \'etoile \`a neutrons (mais qui est
invisible). L'\'etude des instants successifs d'arriv\'ee des pulses
radio a permis de prouver que la p\'eriode orbitale du pulsar autour
de son compagnon d\'ecro\^it tr\`es l\'eg\`erement au cours du temps,
et que cet effet est d\^u \`a l'\'emission du rayonnement gravitationnel
(voir la figure~\ref{fig1}). Plusieurs autres pulsars binaires ont
\'et\'e d\'ecouverts depuis, notamment un syst\`eme pour lequel les
deux \'etoiles \`a neutrons sont simultan\'ement vues comme pulsars,
et qui fournit maintenant le test le plus pr\'ecis sur les ondes
gravitationnelles. Mais, \`a ce jour, le Graal des observations de
pulsars, qui serait un pulsar en orbite autour d'un trou noir, et
permettrait des tests tr\`es int\'eressants de la relativit\'e
g\'en\'erale, n'a pas encore \'et\'e d\'ecouvert.

\vspace{0.3cm}
\section{La d\'etection directe des ondes gravitationnelles sur Terre}
\vspace{0.3cm}

\begin{figure}[t]
\begin{center}
\includegraphics[width=17cm]{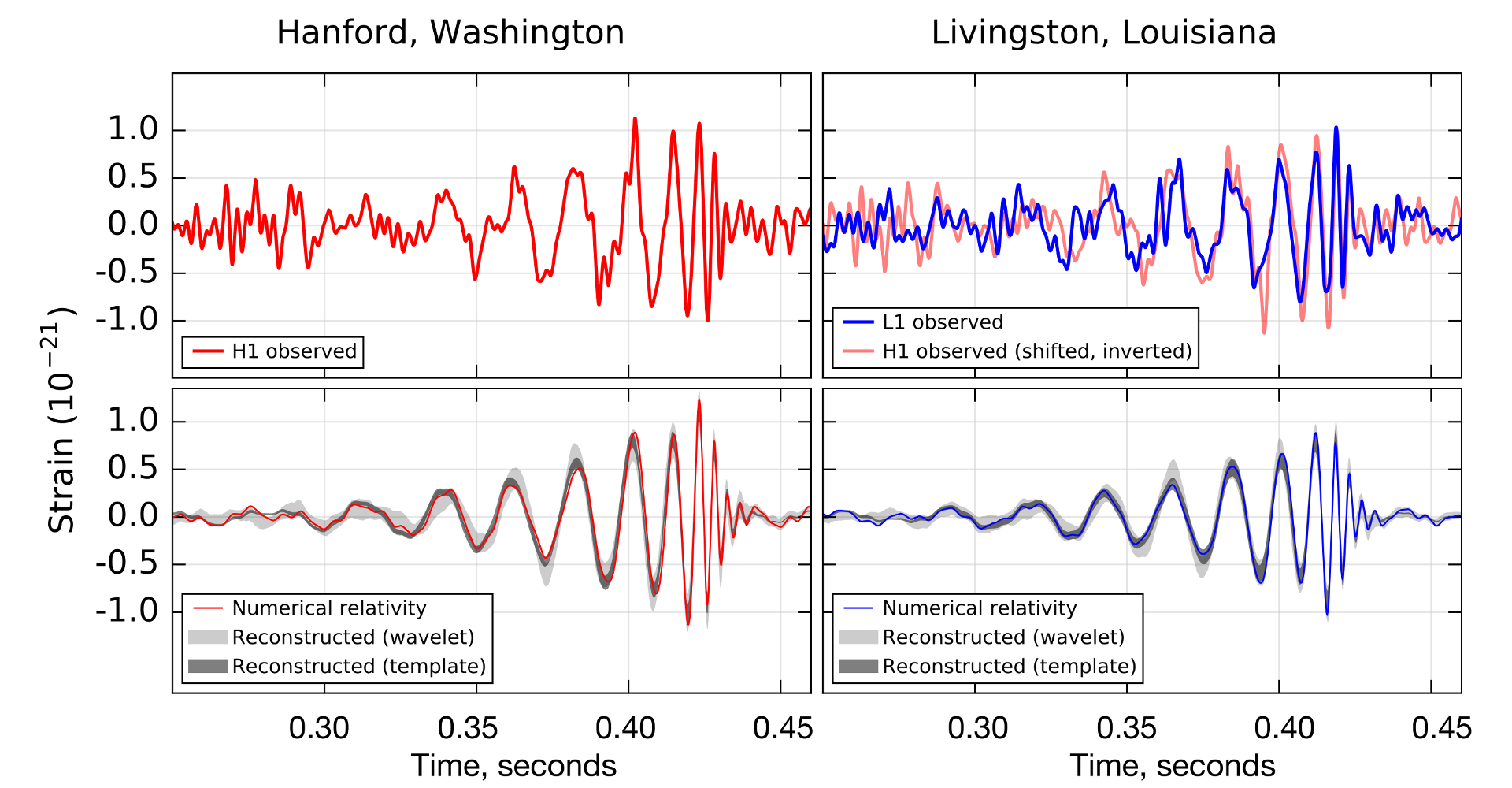}
\end{center}
\caption{En haut, les signaux observ\'es par les d\'etecteurs LIGO
  situ\'es \`a Hanford (en rouge) et Livingstone (en bleu), o\`u
  l'ordonn\'ee repr\'esente l'amplitude de la d\'eformation induite
  sur le d\'etecteur par l'onde gravitationnelle, en unit\'e de
  $10^{-21}$. Les deux signaux se superposent parfaitement apr\`es
  translation en temps de 7 millisecondes. En bas, les meilleurs
  ajustements des signaux observ\'es par les r\'esultats de calculs
  num\'eriques en relativit\'e g\'en\'erale, et par une forme d'onde
  bas\'ee sur une superposition d'ondelettes (mais non physique, car
  ne provenant pas d'un mod\`ele physique).}\label{fig2}
\end{figure}

Le 14 septembre 2015 les d\'etecteurs de la collaboration LIGO-VIRGO
(voir l'article exp\'erimental associ\'e) observ\`erent le signal
d'une onde gravitationnelle produite lors de la collision et la fusion
de deux trous noirs~\cite{GW150914}. L'onde arriva avec un \'ecart
d'environ 7 millisecondes sur les deux d\'etecteurs (Hanford et
Livingston aux Etats-Unis) qui sont s\'epar\'es de 3000 km soit
environ 10 millisecondes-lumi\`ere, ce qui donne une information sur
la direction de la source du signal gravitationnel. Cet \'ev\`enement
appel\'e GW150914 (pour ``Gravitational Wave'' suivit de la date)
s'est produit \`a une distance de environ 400 Mpc,\footnote{Le parsec
  (pc) est la distance \`a laquelle l'unit\'e astronomique (UA,
  distance de la Terre au Soleil) sous-tend un angle d'une seconde
  d'arc, et vaut 3,26 ann\'ees-lumi\`ere (al).} soit 1,3 milliards
d'ann\'ees-lumi\`ere. Donc l'onde gravitationnelle s'est propag\'ee
pendant 1,3 milliards d'ann\'ees jusqu'\`a nous, o\`u elle a
provoqu\'e sur Terre une infime vibration de l'espace-temps mise en
\'evidence par les d\'etecteurs. Le 26 d\'ecembre 2015 \'etait
d\'etect\'e un autre \'ev\'enement de fusion de deux trous noirs
(appel\'e GW151226), situ\'e \`a une distance comparable mais moins
puissant car les trous noirs avaient des masses plus faibles. Un
troisi\`eme \'ev\'enement est suspect\'e mais \'etant beaucoup plus
faible n'a pas pu \^etre confirm\'e, et il est appel\'e plus
modestement LVT151012 (pour ``LIGO-VIRGO trigger''). La prouesse
exp\'erimentale de cette d\'etection est expliqu\'ee dans l'article
exp\'erimental associ\'e. La figure~\ref{fig2} montre le signal
observ\'e de l'\'ev\`enement GW150914.

La premi\`ere constatation que l'on peut faire sur l'interpr\'etation
physique du signal comme d\^u \`a une coalescence est que la formule du
quadrup\^ole d'Einstein de 1918 marche\,! D'apr\`es cette formule
l'\'evolution de la fr\'equence de l'onde gravitationnelle en fonction
du temps dans la phase pr\'ec\'edant la coalescence des deux trous
noirs d\'epend d'une combinaison des deux masses $m_1$ et $m_2$ dite
masse de ``chirp'' ou de ``gazouillement'', qui est donn\'ee par
$\mathscr{M} = \mu^{3/5} M^{2/5}$ o\`u $M = m_1+m_2$ est la masse
totale et $\mu = m_1 m_2/M$ est la masse r\'eduite. Cette appellation
imag\'ee illustre le fait que la fr\'equence du signal, qui est
\'egale \`a deux fois la fr\'equence orbitale, augmente au cours du
temps (\`a cause de la perte d'\'energie li\'ee \`a l'\'emission de
l'onde gravitationnelle), ce qui fait que le signal traduit en onde
sonore ressemble au gazouillement d'un oiseau.

Avec la fr\'equence observ\'ee $f$ et sa variation $\ud f/\ud t$ on
peut donc mesurer $\mathscr{M}$ par la formule du quadrup\^ole et on
trouve environ 30 fois la masse du soleil $M_\odot$ dans le cas de
GW150914, ce qui implique que la masse totale $M$ doit \^etre de
l'ordre de ou sup\'erieure \`a 70 $M_\odot$\,! Avec une analyse plus
fine du signal, utilisant une formule post-newtonienne qui permet de
mesurer les deux masses s\'epar\'ement, on trouve $m_1 = 36 \,M_\odot$
et $m_2 = 29 \,M_\odot$. On est donc en pr\'esence de deux trous noirs
tr\`es massifs. D'autre part, la masse du trou noir final form\'e par
la coalescence est mesur\'ee \`a $M_\text{f} = 62 \,M_\odot$ gr\^ace \`a
une comparaison avec un calcul num\'erique.

La masse de gazouillement $\mathcal{M}$ \'etant connue, on peut alors
mesurer avec l'amplitude du signal la distance \`a laquelle a eu lieu
la coalescence, soit $D = 400 \,\text{Mpc}$ pour GW150914. Une telle
distance correspond \`a l'\'echelle des grandes structures dans
l'Univers, dont on sait depuis une vingtaine d'ann\'ees qu'elles
forment une gigantesque ``toile d'araign\'ee'' avec un r\'eseau de
filaments compos\'es de milliers de galaxies, et joints entre eux par
des super-amas de galaxies et de gaz chaud intergalactique. Mais, d'un
point de vue cosmologique, cette distance se situe dans l'Univers
proche, car le redshift cosmologique\footnote{Le redshift $z$ d'une
  galaxie ou d'une supernova est une mesure de distance d\'efinie par
  le d\'ecalage relatif entre les longueurs d'onde de photons
  re\c{c}us et \'emis, cons\'equence directe de l'expansion de
  l'Univers.} associ\'e est faible, de l'ordre de $z=0,1$.

L'\'energie \'emise par la coalescence sous forme d'ondes
gravitationnelles peut aussi \^etre grossi\`erement estim\'ee par la
formule du quadrup\^ole. Dans le cas de GW150914, cette \'energie
correspond \`a la conversion de $3 \,M_\odot$ en \'energie
gravitationnelle, puisque c'est la diff\'erence entre la masse
initiale $m_1+m_2 = 65 \,M_\odot$ et la masse finale $M_\text{f} = 62
\,M_\odot$. L'\'energie gravitationnelle est \'egale \`a celle qui
serait \'emise en rayonnement \'electromagn\'etique par plusieurs
milliers de supernovas, et, de plus, cette \'energie est \'emise en
quelques dizi\`emes de secondes, ce qui correspond \`a une puissance
colossale de environ $10^{49} \,\text{W}$\,!  N\'eanmoins, cette
puissance est faible si on la compare \`a la puissance maximale \`a
laquelle on s'attend dans une th\'eorie de gravitation
relativiste. Celle-ci correspond \`a la limite d'une source
ultra-compacte et ultra-relativiste, et qui \'emettrait le rayonnement
avec un rendement maximal. Elle est form\'ee uniquement des constantes
fondamentales, soit $c$ la vitesse de la lumi\`ere et $G$ la constante
de Newton, et vaut $\mathscr{P}_\text{max} = c^5/G$ (ce qui n'est
autre que l'unit\'e de Planck pour une puissance, qui se trouve ne pas
d\'ependre de la constante de Planck $\hbar$). En effet,
$\mathscr{P}_\text{max} = 3,63\,10^{52} \,\text{W}$ et on pourrait
imaginer d\'ecouvrir d'autres types de sources encore plus puissantes
que la coalescence des syst\`emes binaires de trous noirs
(l'astronomie gravitationnelle pourrait nous surprendre)\,!

D'o\`u proviennent ces $3 \,M_\odot$ converties en \'energie
gravitationnelle\,? En relativit\'e g\'en\'erale la masse-\'energie
totale associ\'ee \`a un espace-temps est une constante, on l'appelle
la masse ADM (pour ``Arnowitt-Deser-Misner''). Dans le cas de GW150914
cette masse se compose de la somme des masses des deux trous noirs ($M
= m_1+m_2$), plus l'\'energie de liaison associ\'ee \`a leur
interaction gravitationnelle, soit $-G m_1 m_2/2r$ (dans un mod\`ele
newtonien simple o\`u on suppose une orbite circulaire), plus
l'\'energie dans le champ d'ondes gravitationnelles, \'egale \`a
l'int\'egrale de $-\infty$ dans le pass\'e jusqu'\`a l'instant
pr\'esent du flux d'\'energie. En premi\`ere approximation le flux
d'\'energie est donn\'e par la formule du quadrup\^ole. Par
conservation de l'\'energie ADM on voit que l'\'energie dans le champ
d'ondes gravitationnelles est \'egale \`a l'\'energie de liaison
gravitationnelle des trous noirs au moment de la fusion, donc pour une
s\'eparation $r$ de l'ordre de $2GM/c^2$. C'est donc cette \'energie
de liaison qui a \'et\'e prise sur le syst\`eme binaire et emport\'ee
par l'onde gravitationnelle.

\vspace{0.3cm}
\section{Patrons d'ondes gravitationnelles et probl\`eme des deux corps en relativit\'e g\'en\'erale}
\vspace{0.3cm}

Avec un fort rapport signal-sur-bruit et une fr\'equence au moment de
la fusion proche du maximum de sensibilit\'e des d\'etecteurs,
GW150914 est un \'ev\`enement magnifique qui permet de tester la
relativit\'e g\'en\'erale dans un r\'egime de champ gravitationnel
fort et rapidement variable. Le deuxi\`eme \'ev\'enement GW151226 est
aussi tr\`es int\'eressant, car, \'etant donn\'e que les masses sont
plus faibles, il donne acc\`es \`a la phase pr\'ec\'edant
imm\'ediatement la fusion finale, pendant laquelle les deux trous
noirs d\'ecrivent une orbite ``spiralante'', avec une cinquantaine de
cycles orbitaux mesur\'es dans le cas de GW151226.

Pour l'interpr\'etation des signaux observ\'es le probl\`eme
th\'eorique \`a r\'esoudre est celui du mouvement et du rayonnement
des deux corps sous l'action de forces purement gravitationnelles. En
effet, on peut montrer que dans le cas d'objets compacts comme les
trous noirs ou les \'etoiles \`a neutrons, les effets non
gravitationnels, tels que la pr\'esence d'un disque de mati\`ere
autour des \'etoiles, ou les effets du milieu interstellaire et des
champs magn\'etiques, ou m\^eme les effets de structure interne dans
le cas d'\'etoiles \`a neutrons, jouent un r\^ole
n\'egligeable. Essentiellement, la dynamique des deux corps compacts
et les ondes gravitationnelles \'emises d\'ependent uniquement de
leurs masses $m_1$ et $m_2$, et, \'eventuellement, de leurs spins
$S_1$ et $S_2$. Par contre, la mod\'elisation de l'onde
gravitationnelle va bien entendu requ\'erir et d\'ependre d'une
th\'eorie de la gravitation, qui sera soit la relativit\'e
g\'en\'erale soit une th\'eorie alternative. De nos jours on dispose
de tout un arsenal de th\'eories alternatives de la gravitation, qui
vont pouvoir \^etre test\'ees avec les ondes gravitationnelles.

\begin{figure}[t]
\begin{center}
\includegraphics[width=18cm]{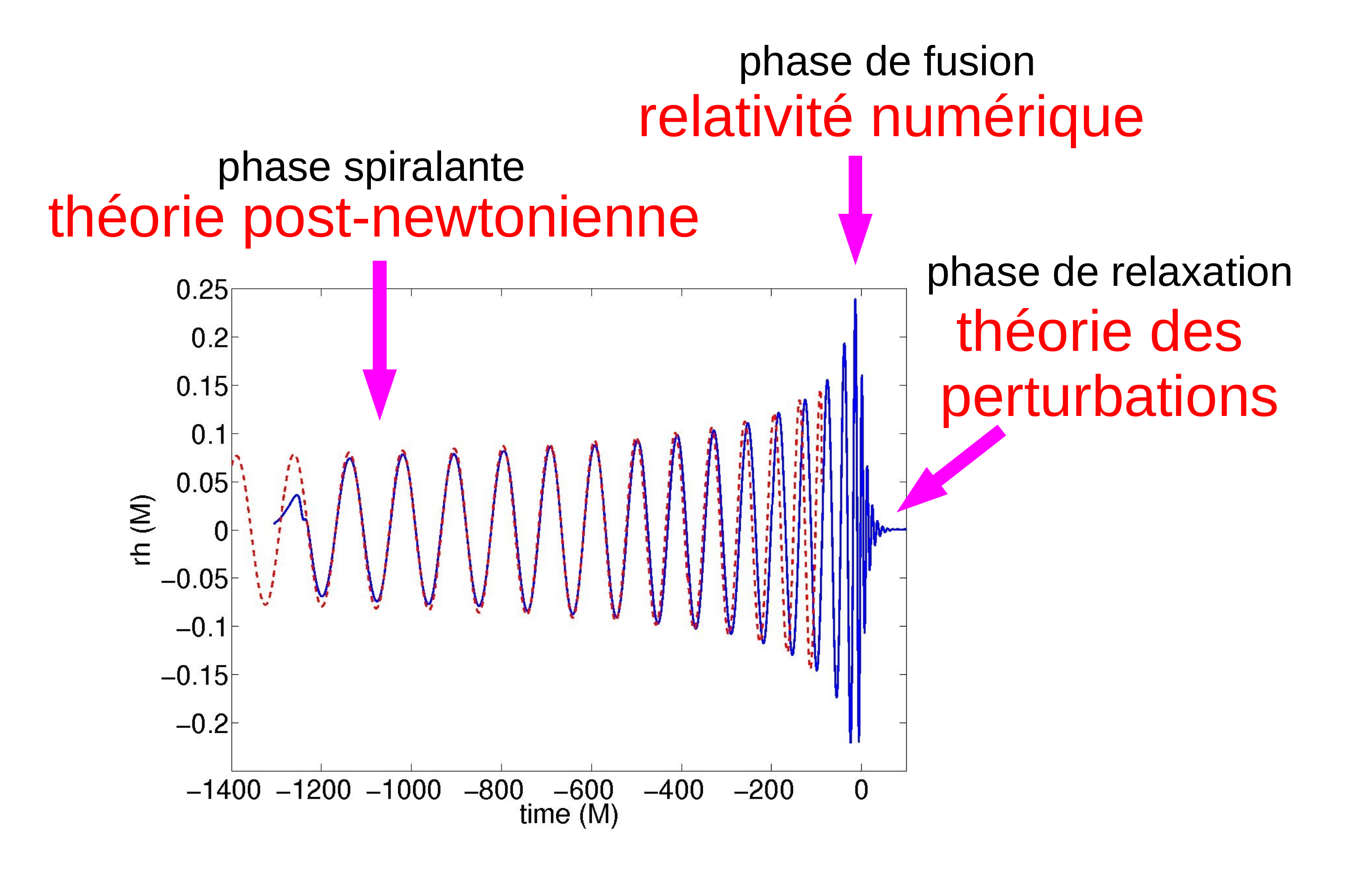}
\end{center}
\caption{L'onde gravitationnelle dans les trois phases de la
  coalescence d'un syst\`eme binaire d'objets compacts, avec les
  m\'ethodes utilis\'ees pour la d\'eterminer. Le calcul analytique
  post-newtonien valable dans la phase spiralante est repr\'esent\'e
  en pointill\'es rouges. Il se raccorde avec le calcul num\'erique
  valable dans les phases de fusion et de relaxation et qui est
  repr\'esent\'e en trait bleu. La th\'eorie des perturbations de
  trous noirs est aussi utilis\'ee pour caract\'eriser les modes de
  vibrations du trou noir final. La dur\'ee relative des trois phases,
  telles que vues dans la bande de fr\'equence d'un d\'etecteur
  donn\'e, d\'epend de la valeur des masses. }\label{fig3}
\end{figure}

On distingue g\'en\'eralement trois phases successives dans le
processus de coalescence de deux trous noirs en relativit\'e
g\'en\'erale (voir la figure~\ref{fig3})\,:
\begin{enumerate}
\item La phase spiralante initiale pendant laquelle la fr\'equence
  orbitale et l'amplitude du signal croissent de fa\c{c}on adiabatique
  \`a cause de la perte d'\'energie li\'ee \`a l'\'emission du
  rayonnement gravitationnel. Dans cette phase la binaire compacte est
  mod\'elis\'ee avec grande pr\'ecision par un syst\`eme de deux
  masses ponctuelles $m_1$ et $m_2$ (avec spins $S_1$ et $S_2$),
  gr\^ace \`a l'approximation post-newtonienne (PN) de la relativit\'e
  g\'en\'erale, qui est un d\'eveloppement lorsque la vitesse des
  corps est faible par rapport \`a la vitesse de la lumi\`ere, donc
  quand le rapport $v/c$ tend formellement vers
  z\'ero~\cite{Bliving14}. La forme d'onde dans la phase spiralante
  est actuellement connue gr\^ace \`a des travaux ayant d\'ebut\'e
  dans les ann\'ees 1980 jusqu'\`a l'ordre post-newtonien tr\`es
  \'elev\'e dit 3,5PN, correspondant \`a des corrections relativistes
  d'ordre $(v/c)^7$ au del\`a de la formule du quadrup\^ole
  d'Einstein.\footnote{La terminologie post-newtonienne a \'et\'e
    introduite par Chandrasekhar dans les ann\'ees 1960. Un terme
    $n$PN (o\`u $n$ est entier ou demi entier) est par d\'efinition
    d'ordre $(v/c)^{2n}$, de sorte que l'on appelle 1PN la premi\`ere
    correction post-newtonienne qui est d'ordre $(v/c)^2$.}
\item La phase de fusion o\`u les horizons des deux trous noirs entrent
  en contact et forment un trou noir unique. La m\'ethode
  post-newtonienne n'est plus valable dans cette phase qui doit \^etre
  trait\'ee par une int\'egration num\'erique des \'equations
  d'Einstein. Pendant longtemps le calcul num\'erique a constitu\'e un
  d\'efi pour la communaut\'e (on l'appelait le ``binary black hole
  challenge''), mais a finalement abouti en 2005~\cite{Pret05}, et
  nous disposons donc maintenant de la solution ``exacte'' (quoique
  num\'erique) pour la phase de fusion. Cette solution est raccord\'ee
  avec grande pr\'ecision \`a la solution post-newtonienne dans la phase
  pr\'ec\'edente.
\item La phase finale dite de relaxation. Le trou noir engendr\'e par
  la fusion est initialement d\'eform\'e \`a cause de la dynamique de
  la collision et \'emet des ondes gravitationnelles correspondant \`a
  ses modes dits ``quasi-normaux'', qui sont les modes de vibration
  intrins\`eques du trou noir, pour finalement se relaxer vers un
  r\'egime stationnaire, qui n'\'emet donc plus de rayonnement et est
  d\'ecrit par le trou noir de Kerr. Cette phase est tr\`es
  int\'eressante pour effectuer des tests des th\'eor\`emes sur les
  trous noirs\,: le trou noir form\'e est-il bien donn\'e par la
  solution du trou noir en rotation de Kerr en relativit\'e
  g\'en\'erale, caract\'eris\'e uniquement par sa masse et son spin\,?
\end{enumerate}

Outre les m\'ethodes ``fondamentales'' que sont l'approximation
post-newtonienne et la relativit\'e num\'erique, des m\'ethodes
analytiques ``effectives'' ont aussi \'et\'e d\'evelopp\'ees, avec
l'ambition de d\'ecrire en une seule fois les trois phases de la
coalescence. Une de ces m\'ethodes consiste \`a partir du probl\`eme
physique r\'eel \`a deux corps d\'ecrit par la solution
post-newtonienne, et \`a le transformer en un probl\`eme plus simple dit
``effectif \`a un corps'' (``effective-one-body'' ou
EOB)~\cite{BuonD99}. L'analyse des donn\'ees dans les d\'etecteurs
LIGO-VIRGO exploite ce type de solution effective, tr\`es utile en
pratique car elle permet de rechercher les signaux en temps r\'eel
dans une grande r\'egion de l'espace des param\`etres (masses et spins
des trous noirs).

Les ondes gravitationnelles d\'ecouvertes par la collaboration
LIGO-VIRGO sont parfaitement en accord avec la pr\'ediction de la
relativit\'e g\'en\'erale dans les trois phases successives de la
coalescence, c'est-\`a-dire que les oscillations observ\'ees dans la
figure~\ref{fig2} sont bien compatibles avec celles pr\'edites dans la
figure~\ref{fig3}\,!  Aucune d\'eviation par rapport \`a la
relativit\'e g\'en\'erale n'a \'et\'e mise en \'evidence au niveau de
la pr\'ecision de la mesure, et la coh\'erence des signaux observ\'es
avec la relativit\'e g\'en\'erale est remarquable.

\begin{figure}[t]
\begin{center}
\includegraphics[width=15cm]{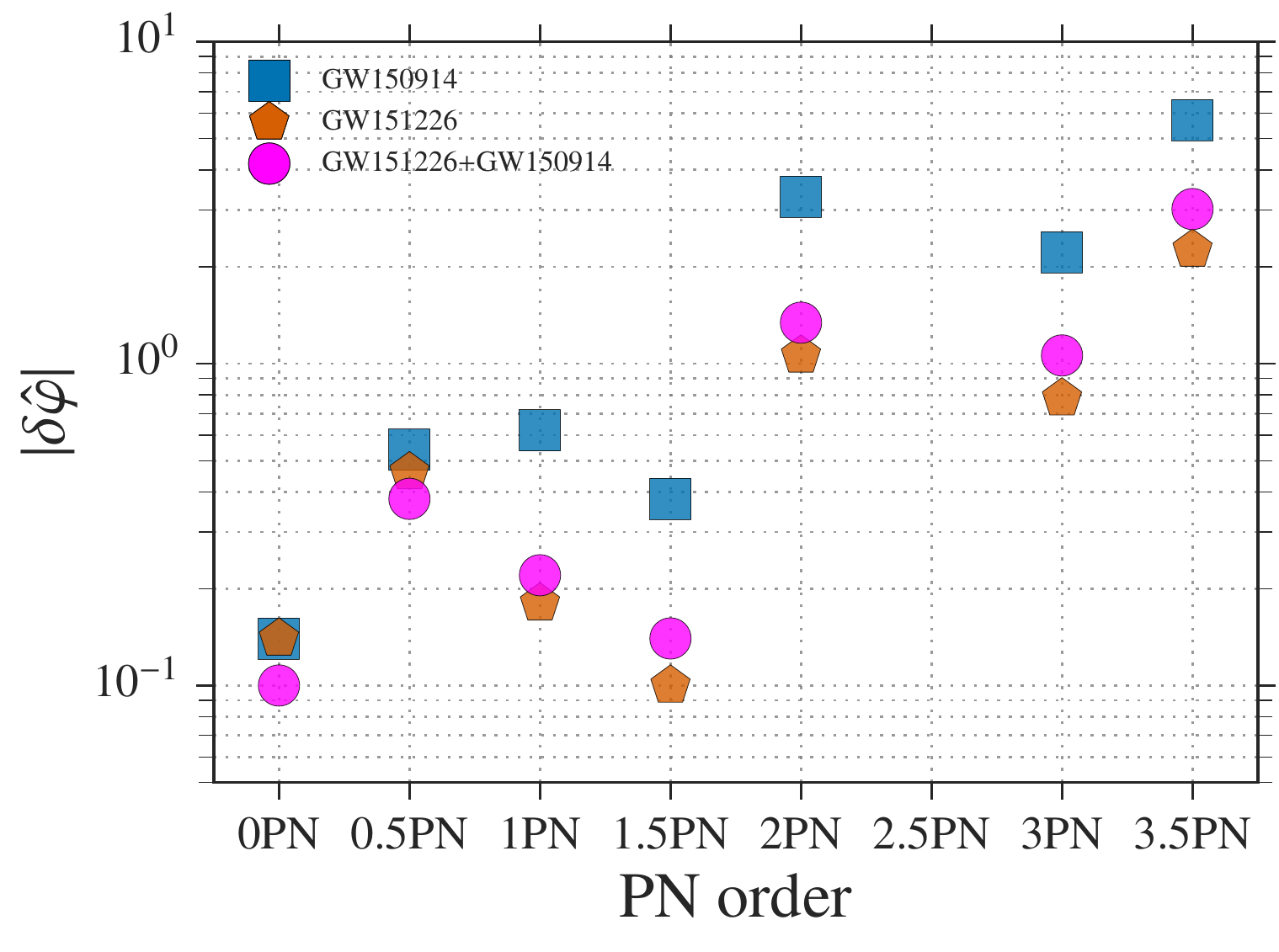}
\end{center}
\caption{Limites sup\'erieures sur les d\'eviations \`a la pr\'ediction
  de la relativit\'e g\'en\'erale pour les param\`etres
  post-newtoniens (qui sont connus jusqu'\`a l'ordre 3,5PN) en utilisant
  les observations combin\'ees de GW150914 et
  GW151226~\cite{LIGOrun1}.}\label{fig4}
\end{figure}

Par exemple, les param\`etres post-newtoniens pr\'evus par la
relativit\'e g\'en\'erale (et qui repr\'esentent des corrections
relativistes \`a la formule du quadrup\^ole) sont d\'ej\`a test\'es
avec les observations de LIGO-VIRGO, surtout gr\^ace \`a
l'\'ev\`enement GW151226 qui donne acc\`es \`a une certaine portion de
la phase spiralante, voir la figure~\ref{fig4}. Ces param\`etres sont
importants car ils sondent la structure non-lin\'eaire de la
relativit\'e g\'en\'erale. Ainsi, \`a partir de l'ordre 1,5PN ou
$(v/c)^3$ dans le d\'eveloppement post-newtonien, apparaissent les
``sillages'' d'ondes gravitationnelles, qui sont des effets
non-lin\'eaires d\^us physiquement au fait que les ondes se propagent
dans un espace-temps qui est lui-m\^eme d\'eform\'e par la pr\'esence
de la source. La mesure effectu\'ee du param\`etre 1,5PN (\`a environ
10\% pr\`es, voir la figure~\ref{fig4}) est une v\'erification de la
pr\'esence de cet effet dans le champ d'ondes gravitationnelles. C'est
un test fin de la relativit\'e g\'en\'erale qu'il est impossible de
r\'ealiser dans le syst\`eme solaire ni m\^eme avec les pulsars
binaires\,!

\vspace{0.3cm}
\section{Quelques r\'eflexions sur l'astronomie gravitationnelle}
\vspace{0.3cm}

L'astronomie gravitationnelle qui d\'emarre avec cette d\'etection
nous laisse entrevoir d'\'enormes avanc\'ees tant en Astrophysique
qu'en Physique Fondamentale. On s'attend par exemple \`a ce que les
ondes gravitationnelles se r\'ev\`elent \^etre une mine d'informations
sur l'\'etat de l'Univers \`a des tr\`es grandes distances, et donc \`a
des instants tr\`es recul\'es dans le pass\'e, jusqu'aux premiers
instants juste apr\`es le Big Bang.

Les astronomes sont maintenant des explorateurs qui peuvent non
seulement voir l'Univers mais aussi l'``\'ecouter''\,!  L'astronomie
gravitationnelle est donc tr\`es compl\'ementaire de l'astronomie
\'electromagn\'etique, en permettant d'\'ecouter des \'ev\`enements
violents survenant dans l'Univers, et qui sont en g\'en\'eral
invisibles dans toute autre forme de rayonnement. Mais, pour certaines
sources, on s'attend \`a la d\'etection concomitante avec les ondes
gravitationnelles de signaux \'electromagn\'etiques ou de particules
\'energ\'etiques comme les neutrinos. Cette astronomie
``multi-messag\`ere'' devrait constituer un \'el\'ement essentiel de
l'exploration de l'Univers gravitationnel et va aussi enrichir les
secteurs de l'astronomie traditionnelle. Par exemple, la d\'etection
de l'onde gravitationnelle en provenance de la coalescence de deux
\'etoiles \`a neutrons pourrait s'accompagner de l'observation d'un
sursaut \'electromagn\'etique $\gamma$ ``court'' (par opposition aux
sursauts ``longs'' qui ont probablement une autre origine), car le
meilleur mod\`ele actuel pour le ``moteur central'' de la production
des sursauts $\gamma$ courts est la coalescence de deux objets
compacts.

Comme en t\'emoigne la mesure des param\`etres post-newtoniens
(figure~\ref{fig4}), l'astronomie gravitationnelle va \^etre une
astronomie de pr\'ecision. Elle nous a d\'ej\`a apport\'e, outre le
fait que la relativit\'e g\'en\'erale reste correcte dans un r\'egime
de champs gravitationnels forts, et que les m\'ethodes d'approximation
``marchent'', la premi\`ere preuve exp\'erimentale directe de
l'existence des trous noirs\,! Mais d\'ej\`a de nombreuses questions
se posent.

\begin{figure}[t]
\begin{center}
\includegraphics[width=14cm]{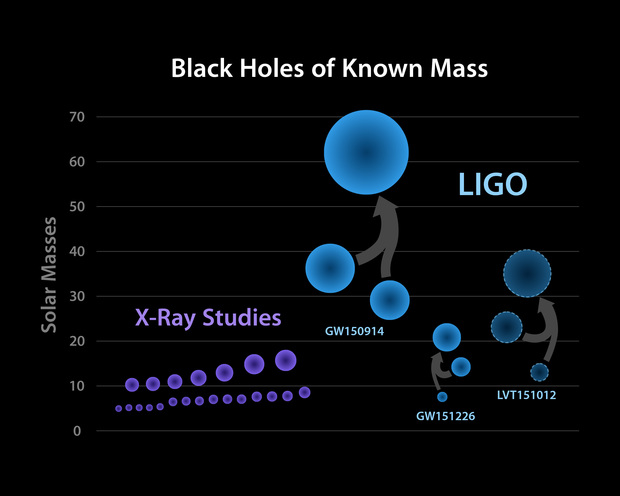}
\end{center}
\caption{Les masses des trous noirs connus dans notre galaxie et
  d'origine stellaire (c'est-\`a-dire issus de l'explosion d'une
  supernova), observ\'es gr\^ace aux rayons X provenant d'un disque
  d'accr\'etion de mati\`ere d\'evers\'ee par un compagnon,
  compar\'ees aux masses des trous noirs mesur\'ees dans les trois
  \'ev\`enements d'ondes gravitationnelles.}\label{fig5}
\end{figure}

Par exemple les masses des trous noirs en jeu dans l'\'ev\`enement
GW150914 sont tr\`es sup\'erieures \`a celles des trous noirs connus
d'origine stellaire dans notre galaxie, voir la
figure~\ref{fig5}. Comment de tels trous noirs tr\`es massifs ont-ils
pu se former\,? La masse d'un trou noir issu de l'explosion en
supernova d'une \'etoile massive d\'epend d'un certain nombres de
param\`etres. L'un des plus importants est la ``m\'etallicit\'e'' de
l'\'etoile prog\'enitrice,\footnote{Pour simplifier on appelle
  ``m\'etal'' tout \'el\'ement dans l'\'etoile qui est plus lourd que
  l'h\'elium.} qui d\'etermine en particulier l'opacit\'e des couches
externes de l'\'etoile. Plus la m\'etallicit\'e est \'elev\'ee et plus
l'\'etoile est opaque \`a la radiation,
ce qui implique une perte de masse importante par vent stellaire, qui
va r\'eduire d'autant la masse du trou noir final form\'e par la
supernova et l'effondrement gravitationnel. Pour former des trous
noirs massifs par explosions de supernovas on a donc besoin d'une
m\'etallicit\'e faible.

Un scenario int\'eressant est de supposer que les trous noirs massifs
ont \'et\'e form\'es dans l'Univers jeune non encore enrichi par les
explosions de supernovas, et donc de m\'etallicit\'e faible. C'est
l'\'epoque dite de la ``r\'eionisation'' o\`u les premi\`eres
\'etoiles apparaissent et r\'eionisent les atomes \`a des redshifts $z
= 10$ \`a $15$, juste apr\`es l'``age sombre'' de la cosmologie. On
imagine que des syst\`emes binaires d'\'etoiles tr\`es massives
\'evoluent et explosent en supernovas, menant \`a la formation d'une
binaire serr\'ee de deux trous noirs massifs. Dans un tel sc\'enario
on a besoin d'invoquer une phase dans l'\'evolution du syst\`eme
binaire pendant laquelle les deux trous noirs se retrouvent entour\'es
d'une enveloppe commune provenant des mat\'eriaux expuls\'es par la
deuxi\`eme explosion de supernova~\cite{Belcz12}. L'enveloppe commune
agit par friction dynamique sur l'orbite des trous noirs et permet de
les rapprocher consid\'erablement, ce qui conduit \`a une paire de
trous noirs assez serr\'ee et qui pourra ensuite \'evoluer par
rayonnement gravitationnel.

\begin{figure}[t]
\begin{center}
\includegraphics[width=14cm]{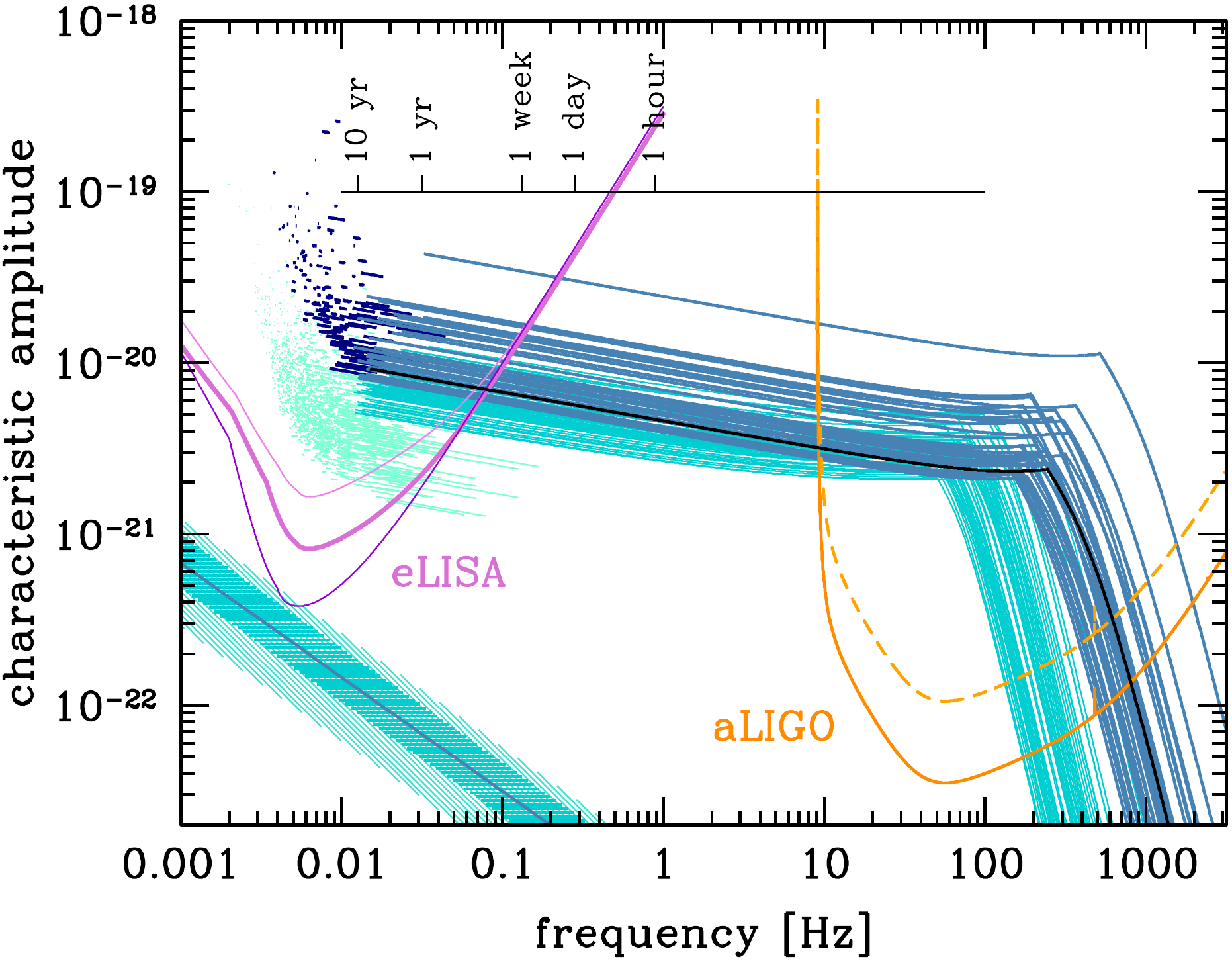}
\end{center}
\caption{L'astronomie multi-longueur d'ondes entre les basses
  fr\'equences d\'etect\'ees dans l'espace par eLISA et les hautes
  fr\'equences d\'etect\'ees au sol par aLIGO (``advanced''
  LIGO-VIRGO). Les lignes oranges indiquent les seuils de
  sensibilit\'e actuel et futur de aLIGO. Les lignes violettes
  indiquent ceux de trois configurations futures de eLISA. Les lignes
  bleues sont les amplitudes des ondes gravitationnelles de syst\`emes
  de trous noirs binaires. L'\'ev\`enement GW150914, indiqu\'e en
  trait plein noir par rapport aux seuils de sensibilit\'es des
  d\'etecteurs, aurait \'et\'e d\'etect\'e environ dix ann\'ees
  auparavant par les d\'etecteurs dans l'espace s'il y en avait eu \`a
  ce moment~\cite{Sesana16}. }\label{fig6}
\end{figure}

Le taux de coalescences de syst\`emes binaires de trous noirs
d'origine stellaire (c'est-\`a-dire issus de l'explosion de
supernovas) est maintenant approximativement connu gr\^ace \`a
GW150914 et GW151226, et se situe entre une dizaine et une centaine
d'\'ev\`enements par an et par Gpc$^3$. Ce taux implique un ciel
gravitationnel tr\`es ``sonore'', non seulement aux fr\'equences
observ\'ees par les d\'etecteurs au sol LIGO-VIRGO, mais aussi \`a des
fr\'equences plus basses accessibles par les futurs observatoires
gravitationnels dans l'espace tels que eLISA (``evolved Laser
Interferometer Space Antenna''). On s'attend ainsi \`a ce que des
milliers de trous noirs binaires d'origine stellaire soient
d\'etectables par eLISA et, parmi eux, des centaines qui fusionneront
plusieurs ann\'ees apr\`es leur d\'etection par eLISA \`a basse
fr\'equence, et seront alors observables \`a haute fr\'equence par les
d\'etecteurs au sol, voir la figure~\ref{fig6}. De plus il devrait
exister un fond stochastique d'ondes gravitationnelles provenant de
syst\`emes binaires de trous noirs stellaires non r\'esolues, qui est
indiqu\'e dans la partie inf\'erieure gauche de la figure~\ref{fig6}.

En conclusion, la d\'ecouverte des ondes gravitationnelles par la
collaboration LIGO-VIRGO marque le d\'emarrage d'une nouvelle \`ere de
l'astronomie moderne, celle de l'exploration de l'Univers
gravitationnel, radicalement diff\'erente et compl\'ementaire de
l'astronomie traditionnelle. L'analyse des d\'etections indique que
cette exploration conduira \`a une nouvelle compr\'ehension de la
structure de l'Univers, notamment \`a l'\'elucidation des m\'ecanismes
de formation des trous noirs et de leur r\^ole dans l'\'evolution de
l'Univers. L'astronomie gravitationnelle va sonder les grandes
\'echelles de distances cosmologiques, et les premiers instants
proches du Big Bang. Elle pourrait s'av\'erer tr\`es riche en
nouvelles d\'ecouvertes, peut-\^etre en synergie avec les m\'ethodes
optiques, et peut-\^etre compl\`etement inattendues\,! Par ailleurs,
cette exploration nous autorise \`a envisager des tests in\'edits de
la gravitation et de la relativit\'e g\'en\'erale en tant que
th\'eorie ``classique'', ouvrant la voie \`a une meilleure
connaissance du statut de l'interaction gravitationnelle, avec des
cons\'equences sur une \'eventuelle description quantique de cette
interaction, et de son unification possible avec les autres
interactions fondamentales.

\bibliography{ListeRef.bib}

\begin{thebibliography}{10}
\expandafter\ifx\csname natexlab\endcsname\relax\def\natexlab#1{#1}\fi
\expandafter\ifx\csname bibnamefont\endcsname\relax
  \def\bibnamefont#1{#1}\fi
\expandafter\ifx\csname bibfnamefont\endcsname\relax
  \def\bibfnamefont#1{#1}\fi
\expandafter\ifx\csname citenamefont\endcsname\relax
  \def\citenamefont#1{#1}\fi
\expandafter\ifx\csname url\endcsname\relax
  \def\url#1{\texttt{#1}}\fi
\expandafter\ifx\csname urlprefix\endcsname\relax\def\urlprefix{URL }\fi
\providecommand{\bibinfo}[2]{#2}
\providecommand{\eprint}[2][]{\url{#2}}

\bibitem[{\citenamefont{Einstein}(1918)}]{E18}
\bibinfo{author}{\bibfnamefont{A.}~\bibnamefont{Einstein}},
  \bibinfo{journal}{Sitzber. Preuss. Akad. Wiss. Berlin}
  \textbf{\bibinfo{volume}{154}} (\bibinfo{year}{1918}).

\bibitem[{\citenamefont{Hulse and Taylor}(1975)}]{HulseTaylor}
\bibinfo{author}{\bibfnamefont{R.}~\bibnamefont{Hulse}} \bibnamefont{and}
  \bibinfo{author}{\bibfnamefont{J.}~\bibnamefont{Taylor}},
  \bibinfo{journal}{Astrophys. J.} \textbf{\bibinfo{volume}{195}},
  \bibinfo{pages}{L51} (\bibinfo{year}{1975}).

\bibitem[{\citenamefont{Taylor and Weisberg}(1982)}]{TW82}
\bibinfo{author}{\bibfnamefont{J.}~\bibnamefont{Taylor}} \bibnamefont{and}
  \bibinfo{author}{\bibfnamefont{J.}~\bibnamefont{Weisberg}},
  \bibinfo{journal}{Astrophys. J.} \textbf{\bibinfo{volume}{253}},
  \bibinfo{pages}{908} (\bibinfo{year}{1982}).

\bibitem[{\citenamefont{Abbott et~al.}(2016{\natexlab{a}})}]{GW150914}
\bibinfo{author}{\bibfnamefont{B.}~\bibnamefont{Abbott}} \bibnamefont{et~al.}
  (\bibinfo{collaboration}{LIGO Scientific Collaboration and VIRGO
  Collaboration}), \bibinfo{journal}{Phys. Rev. Lett.}
  \textbf{\bibinfo{volume}{116}}, \bibinfo{pages}{061102}
  (\bibinfo{year}{2016}{\natexlab{a}}), \eprint{arXiv:1602.03837 [gr-qc]}.

\bibitem[{\citenamefont{Blanchet}(2014)}]{Bliving14}
\bibinfo{author}{\bibfnamefont{L.}~\bibnamefont{Blanchet}},
  \bibinfo{journal}{Living Rev. Rel.} \textbf{\bibinfo{volume}{17}},
  \bibinfo{pages}{2} (\bibinfo{year}{2014}), \eprint{arXiv:1310.1528 [gr-qc]}.

\bibitem[{\citenamefont{Pretorius}(2005)}]{Pret05}
\bibinfo{author}{\bibfnamefont{F.}~\bibnamefont{Pretorius}},
  \bibinfo{journal}{Phys. Rev. Lett.} \textbf{\bibinfo{volume}{95}},
  \bibinfo{pages}{121101} (\bibinfo{year}{2005}), \eprint{gr-qc/0507014}.

\bibitem[{\citenamefont{Buonanno and Damour}(1999)}]{BuonD99}
\bibinfo{author}{\bibfnamefont{A.}~\bibnamefont{Buonanno}} \bibnamefont{and}
  \bibinfo{author}{\bibfnamefont{T.}~\bibnamefont{Damour}},
  \bibinfo{journal}{Phys. Rev. D} \textbf{\bibinfo{volume}{59}},
  \bibinfo{pages}{084006} (\bibinfo{year}{1999}), \eprint{gr-qc/9811091}.

\bibitem[{\citenamefont{Abbott et~al.}(2016{\natexlab{b}})}]{LIGOrun1}
\bibinfo{author}{\bibfnamefont{B.}~\bibnamefont{Abbott}} \bibnamefont{et~al.}
  (\bibinfo{collaboration}{LIGO Scientific Collaboration and VIRGO
  Collaboration}), \bibinfo{journal}{Phys. Rev. X}
  \textbf{\bibinfo{volume}{6}}, \bibinfo{pages}{041015}
  (\bibinfo{year}{2016}{\natexlab{b}}), \eprint{arXiv:1606.04856 [gr-qc]}.

\bibitem[{\citenamefont{Dominik et~al.}(2012)\citenamefont{Dominik, Belczynski,
  Fryer, Holz, Berti, Bulik, Mandel, and O'Shaughnessy}}]{Belcz12}
\bibinfo{author}{\bibfnamefont{M.}~\bibnamefont{Dominik}},
  \bibinfo{author}{\bibfnamefont{K.}~\bibnamefont{Belczynski}},
  \bibinfo{author}{\bibfnamefont{C.}~\bibnamefont{Fryer}},
  \bibinfo{author}{\bibfnamefont{D.}~\bibnamefont{Holz}},
  \bibinfo{author}{\bibfnamefont{E.}~\bibnamefont{Berti}},
  \bibinfo{author}{\bibfnamefont{T.}~\bibnamefont{Bulik}},
  \bibinfo{author}{\bibfnamefont{I.}~\bibnamefont{Mandel}}, \bibnamefont{and}
  \bibinfo{author}{\bibfnamefont{R.}~\bibnamefont{O'Shaughnessy}},
  \bibinfo{journal}{Astrophys. J.} \textbf{\bibinfo{volume}{759}},
  \bibinfo{pages}{52} (\bibinfo{year}{2012}), \eprint{arXiv:1202.4901
  [astro-ph]}.

\bibitem[{\citenamefont{Sesana}(2016)}]{Sesana16}
\bibinfo{author}{\bibfnamefont{A.}~\bibnamefont{Sesana}},
  \bibinfo{journal}{Phys. Rev. Lett.} \textbf{\bibinfo{volume}{116}},
  \bibinfo{pages}{231102} (\bibinfo{year}{2016}), \eprint{arXiv:1602.06951
  [gr-qc]}.

\end{thebibliography}

\end{document}